\documentclass[%
reprint,	% True two column version
superscriptaddress,
showpacs,preprintnumbers,
amsmath,amssymb,
floatfix,
prl,
]{revtex4-1}
\usepackage{graphicx}% Include figure files
\usepackage{dcolumn}% Align table columns on decimal point
\usepackage{bm}% bold math
\begin{document}

\preprint{APS/123-QED}

\title{Phonon excitation and instabilities in biased graphene nanoconstrictions}%nano-bridges
%\title{Current-induced instabilities in graphene nanoconstrictions}%nano-bridges
%\\Current-induced heating and break-down of graphene nanoconstrictions
\author{Tue Gunst}
\email{Tue.Gunst@nanotech.dtu.dk}
\affiliation{Department of Micro- and Nanotechnology (DTU Nanotech), Center for Nanostructured Graphene (CNG), Technical University of Denmark, DK-2800 Kgs. Lyngby, Denmark}
%\affiliation{Center for Nanostructured Graphene (CNG), Technical University of Denmark, DK-2800 Kgs. Lyngby, Denmark}
\author{Jing-Tao \surname{L\"u}}
\email{jtlu@mail.hust.edu.cn, now at School of Physics, Huazhong University of Science and Technology, Wuhan, China.}
\affiliation{Department of Micro- and Nanotechnology (DTU Nanotech), Center for Nanostructured Graphene (CNG), Technical University of Denmark, DK-2800 Kgs. Lyngby, Denmark}
\affiliation{Niels Bohr Institute, Nano-Science Center, University of Copenhagen, Universitetsparken 5, 2100 Copenhagen {\O}, Denmark}
%\affiliation{School of Physics, Huazhong University of Science and Technology, Wuhan, China}
%\author{Troels Markussen}
%\affiliation{Center for Atomic-scale Materials Design (CAMD), Department of Physics, Technical University of Denmark, DK-2800 Kgs. Lyngby, Denmark}%
\author{Per \surname{Hedeg{\aa}rd}}
\affiliation{Niels Bohr Institute, Nano-Science Center, University of Copenhagen, Universitetsparken 5, 2100 Copenhagen {\O}, Denmark}
\author{Mads Brandbyge}
\affiliation{Department of Micro- and Nanotechnology (DTU Nanotech), Center for Nanostructured Graphene (CNG), Technical University of Denmark, DK-2800 Kgs. Lyngby, Denmark}
%\affiliation{Center for Nanostructured Graphene (CNG), Technical University of Denmark, DK-2800 Kgs. Lyngby, Denmark}

\date{\today}% It is always \today, today,

\begin{abstract}
We calculate the phonons in a graphene nanoconstriction(GNC) in the presence of a high current density.
The Joule-heating, current-induced forces, and coupling to electrode phonons is evaluated using first principles nonequilibrium DFT-NEGF calculations. Close to a resonance in the electronic structure we observe a strongly nonlinear heating with bias and breakdown of the harmonic approximation. This behavior results from negatively damped phonons driven by the current. The effect may limit the stability and capacity of graphene nanoconstrictions to carry high currents.
\end{abstract}
%\pacs{63.22.Rc, 72.80.Vp, 72.10.Di, 46.32.+x}% PACS, the Physics and Astronomy
%63.22.Rc Phonons in graphene. 72.80.Vp Electronic transport in graphene
%Use showkeys class option if keyword
\maketitle
%\tableofcontents

% Overview:\\
%1) Simple derivations of Fnc, JH og Berry.\\
%2) Ab initio electron and phonon transport (importance of adatom, doping, bondcurrents; both for elecs og phs) + inelastic change in bondcurrents!\\
%3) Current-induced modes (Q-factor)! plot modes. Gating!\\
%4) Harmonic temperature distribution/heating.
%5) the importance of the finite bias, damping, Anharmonics...

%\section{Introduction}
%\emph{Introduction --}
%\section{   \label{Transport}}
%For the purpose of predicting the device stability we solve the Langevin equation within the harmonic approximation including the effects of the current.
Graphene has emerged as a highly attractive material for future electronic devices\cite{Novoselov2004,Neto2009}.
It can sustain current densities six orders of magnitude larger than copper and is foreseen to be a versatile material with numerous applications in nanoelectronics, spintronics and nanoelectromechanics\cite{geim_graphene:_2009}. In graphene nanoconstrictions (GNCs) the current is passed through a short ribbon\cite{Nakada1996,Brey2006} at the narrowest point. 
Constrictions and nanoribbons provide semi-conducting interconnects in graphene nano-circuitry \cite{areshkin_building_2007,botello-mendez_quantum_2011}, and is a central building block of graphene based nano-electronics.
Related structures include graphene antidot lattices\cite{bai_graphene_2010,pedersen_graphene_2008}, which can be viewed as a periodic network of constrictions. 
Current state-of-the-art experiments indicate that these may be ``sculpted'' in monolayer graphene with close to atomic precision to a width of a few benzene rings\cite{Xu_controllable_2013}. 

Clearly, for GNCs of this size the current density can locally be very high, and it is important to address 
their stability and performance under bias\cite{borrnert_lattice_2012}.
Experimental results for electron transport\cite{tombros_quantized_2011,darancet_coherent_2009}, local heating by Raman spectroscopy\cite{chae_hot_2009,berciaud_electron_2010,jo_low-frequency_2010}, and infrared emission\cite{freitag_thermal_2010}, have been published for GNCs. 
Recently, it has been argued that several current-induced forces and excitation mechanisms driven by these, besides Joule heating, can play a role for the 
stability of nano-conductors\cite{dundas_current-driven_2009,lu_current-induced_2012,bode_scattering_2011,lu_laserlike_2011}. In particular energy nonconservative "wind"/"waterwheel" forces may transfer energy to the phonons in parallel with the well-known Joule heating.
However, it is not easy to directly infer these mechanisms from experiments in most cases. On the other hand for graphene, the structural response to a high bias can be studied by {\it in situ} transmission electron microscopy, making graphene nanoconductors a good test bed for current-induced phenomena\cite{jia_controlled_2009,barreiro_graphene_2012,engelund_localized_2010}. In particular, a gate-electrode can be used to control the electronic states involved in the transport and thereby the current-induced excitation.

\begin{figure}[!htbp]
\centering
%{\includegraphics[width=0.99\linewidth]{EPSfigures/SystemEigChan.eps}}
%%{\includegraphics[width=0.99\linewidth]{EPSfigures/ConstricTransportSetup.eps}}
%{\includegraphics[width=0.99\linewidth]{EPSfigures/TransmissionWithAndWithoutK.eps}}
%%{\includegraphics[width=0.42\paperwidth]{EPSfigures/Mode14ImagNoPhDamping.eps}}
%%{\includegraphics[width=0.48\linewidth]{EPSfigures/BondCurrent2B.eps}}
%%{\includegraphics[width=0.48\linewidth]{EPSfigures/BondCurrent1B.eps}}
%{\includegraphics[width=0.99\linewidth]{EPSfigures/BondcurrentsCollected2.eps}}
{\includegraphics[width=0.99\linewidth]{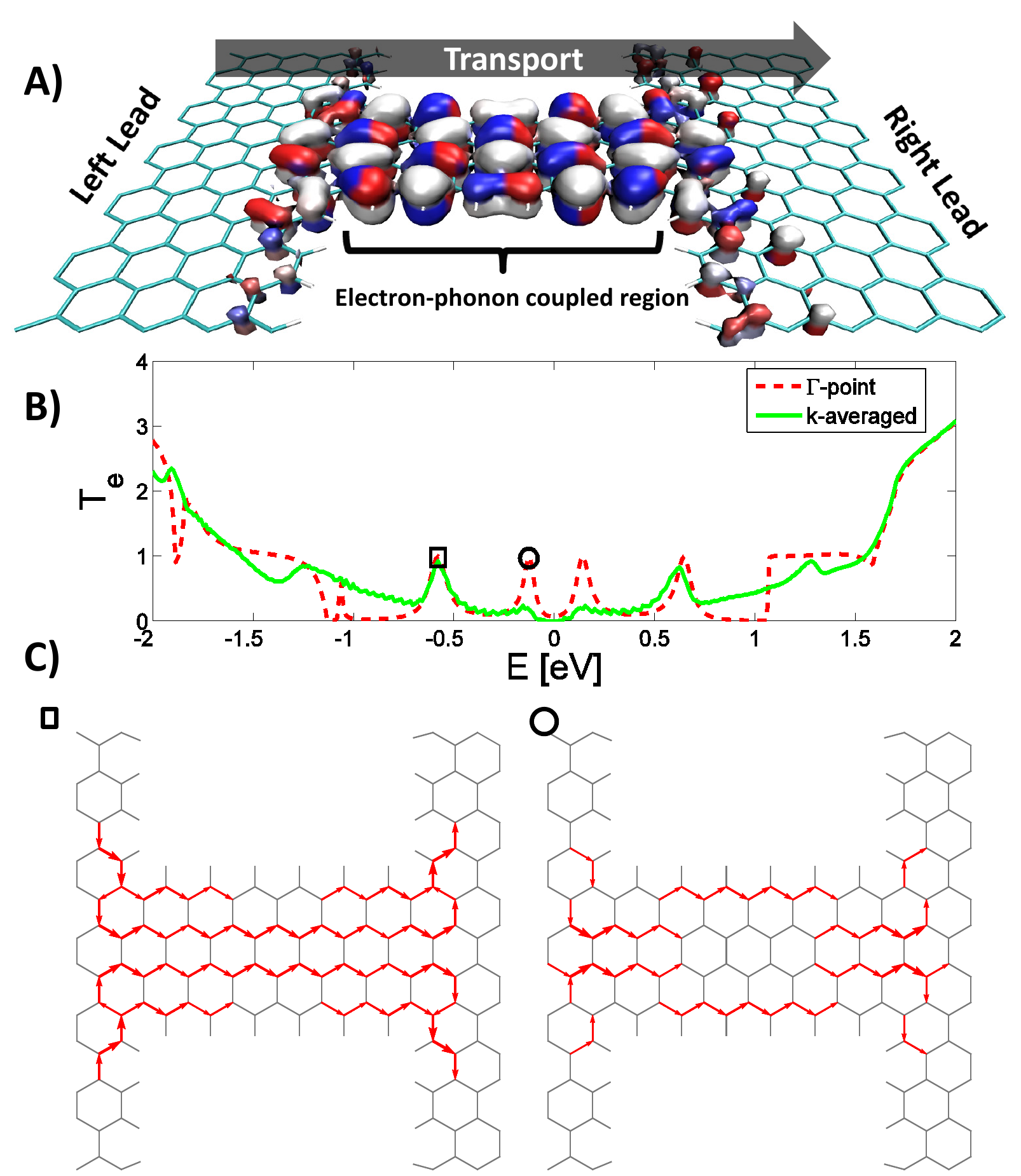}}
\caption{(Color online) A) Transport setup illustrating the hydrogen passivated GNC. The left eigenchannel at zero bias and $E\approx -0.58$eV (colored according to phase, red-white-blue from $-\pi$ to $\pi$). B) $\Gamma$-point and the k-averaged transmission function ($E=0$ corresponds to the Dirac point). C) Bond-currents at the two peaks ($\Gamma$) marked in the transmission plot ($E\approx -0.12$eV and $E\approx -0.58$eV).}
\label{fig:TransmissionAndBondcurrent} %  in a triangular lattice
\end{figure}

% Main conclusions of the paper..
In this Letter, we calculate the current-induced phonon excitation in a small hydrogen-passivated GNC (Fig.~\ref{fig:TransmissionAndBondcurrent}) using parameters obtained from density functional theory (DFT).
We find a highly non-linear heating of the GNC which we trace back  to the deterministic current-induced forces, as opposed to the Joule-heating by random forces. In particular, the nonequilibrium electronic friction force turns into an amplification for certain phonon modes in the GNC. These will dominate the dynamics beyond a certain voltage threshold leading to a breakdown of the harmonic approximation\cite{bode_scattering_2011,lu_laserlike_2011}. Negative friction was theoretically predicted in the tunneling transport through asymmetric molecules\cite{ryndyk_nonequilibrium_2006,lu_laserlike_2011} driven by population inversion between {\em two} molecular states. In contrast, we show here how the GNC can display negative friction due to a build-in asymmetry of phonon emission/absorption. 
%This excitation mechanism, which was proposed for a special molecular electronic structure, is thus applicable to a wider and important class of systems\cite{lu_laserlike_2011}. 

\begin{figure}[htbp]
\centering
%{\includegraphics[width=.5\linewidth]{EPSfigures/TotalPotentialDrop0.5Vv2b.eps}}
%{\includegraphics[width=.48\linewidth]{EPSfigures/IVWithInterPv2c.eps}}
%{\includegraphics[width=0.99\linewidth]{EPSfigures/TransmissionsShiftedV2b.eps}}
{\includegraphics[width=0.99\linewidth]{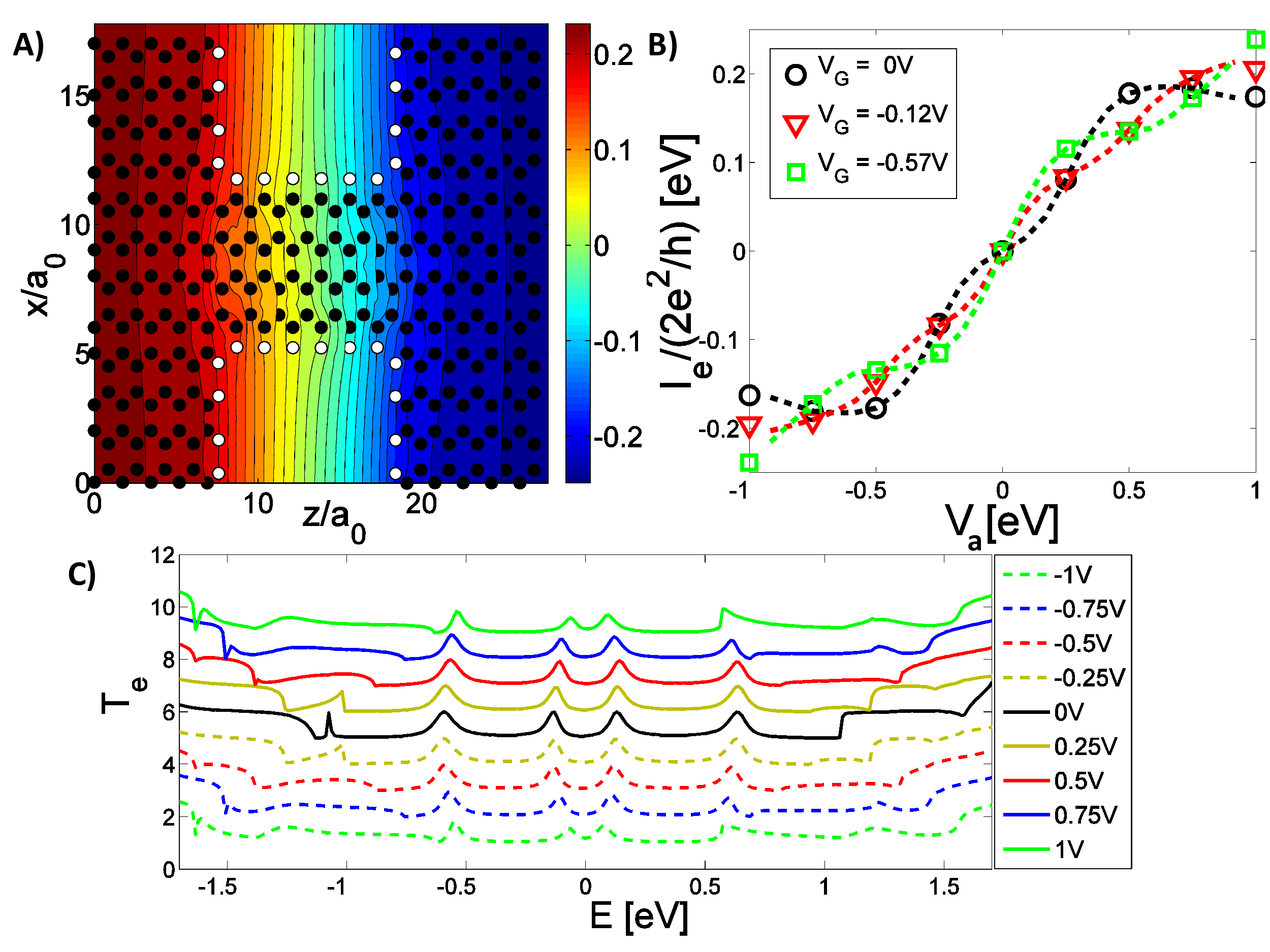}}
\caption{(Color online) A) Real space potential drop ($V_a=0.5$eV) integrated along the out of plane direction (in the region with non-vanishing electronic density). B) $IV$-characteristics for the GNC gated to different chemical potential. Gating to a peak lowers the resistance at low $V_a$. C) Transmission curves (shifted vertically) for different applied bias ($E_F=0$).}
\label{fig:FiniteBias}
\end{figure}

In Fig.~\ref{fig:TransmissionAndBondcurrent}B we see how the 
electron transmission of the GNC for energies around the charge neutral Fermi energy ($E_F=0$) is dominated by two resonance peaks originating from states presenting localized current along the edges(1st peak) and through the center(2nd peak) of the GNC, respectively. Resonances occur due to the diffraction barrier at abrupt interfaces in graphene\cite{darancet_coherent_2009}. By employing a gate voltage($V_g$) we may tune $E_F$ close to a highly conducting peak and consider the phonon excitation close to the resonance.
We will focus on the constriction gated to the 2nd peak, which is mostly unaffected by the boundary conditions in the electrodes(k-point sampling)\cite{brandbyge_density-functional_2002}, and exhibits little dependence on the applied bias ($V_a$), see Fig.~\ref{fig:FiniteBias}.

%The following discussion holds true as long the system is gated to a chemical potential, where the same peak is the first to enter the bias window.

To address the phonon excitation in the presence of current we employ the semi-classical generalized Langevin equation (SCLE)\cite{lu_blowing_2010,bode_scattering_2011,lu_current-induced_2012,lu_current-induced_2011}. The SCLE describe the Joule heating, current-induced forces, and coupling to electrode phonons in the same formalism. For the mass-scaled ion displacements ($Q$) the SCLE reads,
\begin{eqnarray}
\ddot Q(t)&=&-K Q(t)-\int_{}^t \Pi^r(t-t')Q(t')dt' + f_{}(t) \,.
\label{eq:langa}
\end{eqnarray}
Here $K$ is the force constant matrix. The coupling to the electron and phonon baths are described by the retarded phonon self-energies $\Pi^r=\Pi^r_e+\Pi^r_{ph}$, and the random noise force, $f_{}(t)$, accounts for the Joule heating\cite{lu_blowing_2010}. 
We consider the retarded self-energy due to the interaction between the phonons and the electronic current,
\begin{eqnarray}
\Pi_e^r(\omega)  &=& i\pi{\rm Re}\Lambda(\omega) -\pi{\rm Im}\Lambda(\omega) \nonumber\\
&+&\pi\mathcal{H}\{{\rm Re}\Lambda(\omega')\}(\omega)  +i\pi\mathcal{H}\{{\rm Im}\Lambda(\omega')\}(\omega) \,,
\end{eqnarray}
which is given by the interaction-weighted electron-hole pair density of states, $\Lambda$, and its Hilbert transform($\mathcal{H}$)\cite{Hilbert_def}. The four terms in this expression yields the electronic friction, non-conservative wind, renormalization and Berry forces in nonequilibrium conditions, respectively\cite{lu_current-induced_2012}.
Especially for the nonequilibrium electron system,  $\Lambda=\sum_{\alpha,\beta}\Lambda^{\alpha \beta}$, with contributions from left/right leads ($\alpha=L,R$),
\begin{eqnarray}
\Lambda^{\alpha \beta}(\omega) &\equiv& 2 \int \frac{d\epsilon}{4\, \pi^2} {\rm Tr}\left[\mathbf{M}^k \mathbf{A}_{\alpha}(\epsilon+\omega) \mathbf{M}^l \mathbf{A}_{\beta}(\epsilon)\right]\nonumber\\ &\times&\left[n_F(\epsilon+\omega-\mu_{\alpha})-n_F(\epsilon-\mu_{\beta})\right] \,. \label{eq:ephDOS}
\end{eqnarray}
Here $\mathbf{M}^k$ is the coupling to phonon mode $k$, $\mathbf{A}_{\alpha}$ the electronic spectral density for states originating from lead $\alpha$ with chemical potential $\mu_{\alpha}$, and $n_F$ the Fermi distribution.
The spectral density for the noise, $f$, including the Joule heating, is given by,
\begin{eqnarray}
&&S_f(\omega) = -\pi \Lambda(\omega) \coth(\frac{\omega}{2 k_B T})\\
&&- \pi \sum_{\alpha,\beta} \Lambda^{\alpha \beta}(\omega) \left[\coth(\frac{\omega - (\mu_{\alpha}-\mu_{\beta})}{2 k_B T}) - 
\coth(\frac{\omega}{2 k_B T}) \right]\nonumber\,.
\end{eqnarray}
Importantly, we include the full electronic and phononic structure of the graphene electrodes, and go beyond the constant/wide-band approximation(WBA) for the electronic structure. This is essential for our results of the phonon excitation when the graphene system is gated close to electronic resonance. 
We determine all parameters entering the SCLE above in the presence of current using first principles DFT and nonequilibrium Green's functions (DFT-NEGF)\cite{soler_siesta_2002,brandbyge_density-functional_2002,frederiksen_inelastic_2007,FN_settings}. 
We restrict the el-ph interaction to the GNC-region where the current-density is high, and evaluate the electronic spectrum at finite bias, but neglect the small voltage-dependence of $K$ and $\mathbf{M}^k$.

We note that the GNC device-region in the present calculation encompass a basis of 1336 orbitals for the electronic subsystem (matrix size in Eq.~\ref{eq:ephDOS}). Thus in order to efficiently compute $\Lambda$ in Eq.~\ref{eq:ephDOS} beyond WBA we first limited the basis. We employed an expansion of the retarded Green's function and $\mathbf{A}_{\alpha}$ in the eigenspace of ${\bf H}+{\bf \Sigma}_0(E_F)$, ${\bf H}$ being the electronic Hamiltonian and ${\bf \Sigma}_0(E_F)$ the lead self-energies, which vary slowly with energy\cite{egger_vibration-induced_2008}. We have found it sufficient to limit this basis to 200 states within the interval [-7,6]eV around $E_F$. Secondly, we computed $\Lambda$ by parallel execution over the $\omega$ and $V_a$ parameters.

\begin{figure}[htbp]
\centering
%{\includegraphics[width=0.99\linewidth]{EPSfigures/DosRunawayNegBiasWithRNv2b.eps}}
%%{\includegraphics[width=0.99\linewidth]{EPSfigures/EnergySpectrumIntegratedCollectedTotalV2.eps}}
%{\includegraphics[width=0.99\linewidth]{EPSfigures/HeatingNoRNcompareArticleb.eps}}
{\includegraphics[width=0.99\linewidth]{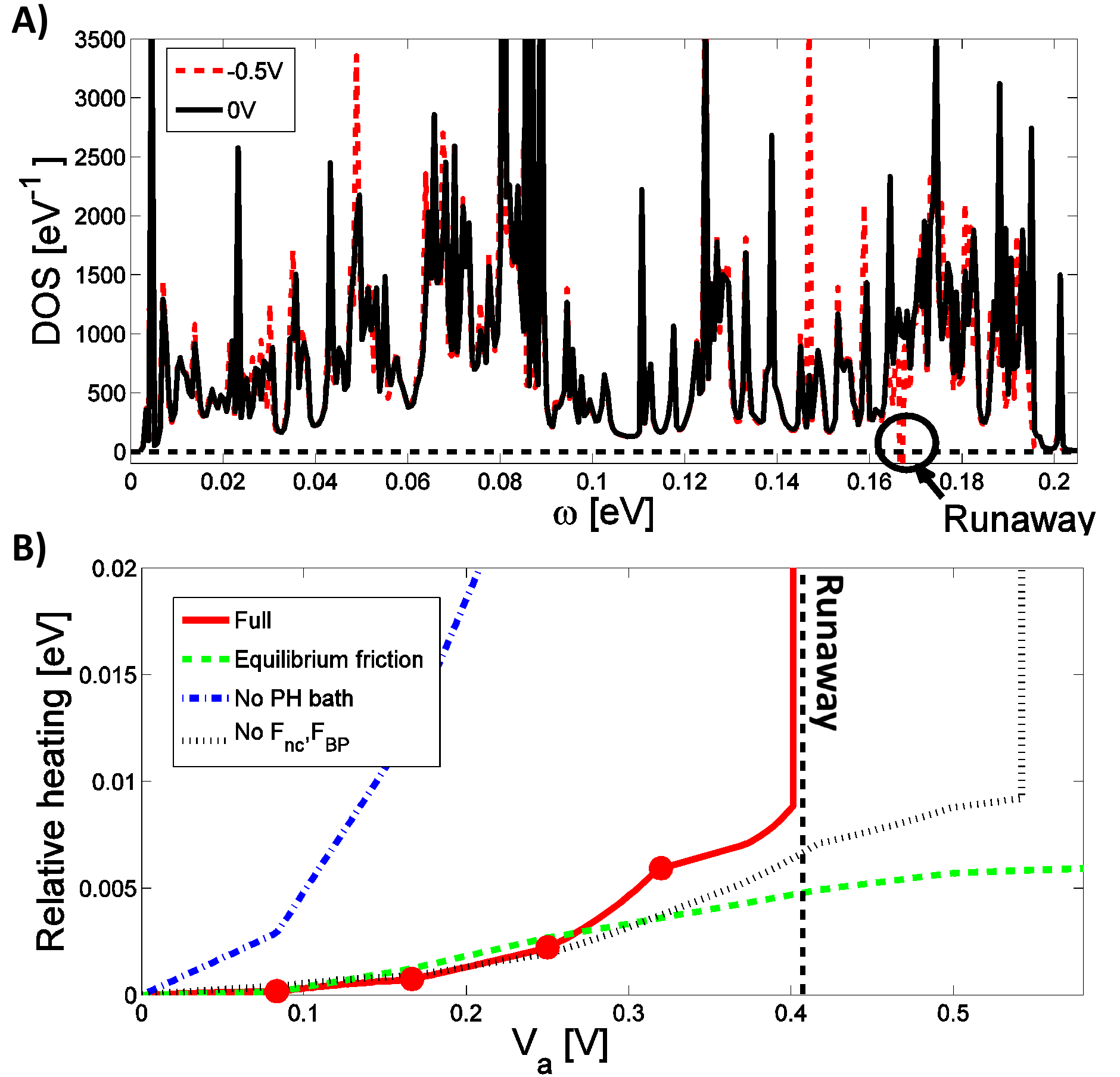}}
\caption{(Color online) A) Dashed(full) lines show the phonon density of states (DOS) of the GNC with(without) electronic current. A unstable "runaway" mode appears for an applied bias of $V_a\approx\pm 0.5$V as a negative DOS peak. 
B) Heating (change in average kinetic energy per atom due to current) of the GNC at $300\,$K. 
Full line: Result incl. all current-induced forces. Dashed line: only fluctuating force (Joule heating) and zero-bias electronic friction. Dot-dashed line: The wide-band approximation(WBA) without coupling to the electrode phonon bath. Dotted line: Full calculation neglecting the wind and Berry-phase forces.}
\label{fig:Heating}%diagonal wide-band.
\end{figure}% The heating is evaluated as the average kinetic energy relative to the zero bias result.

% GF, DOS, Heating (3 contributions = stochastic JH, deterministic friction which is mainly diagonal, off-diagonal nonconservative and Berry phase forces)
% The nonadiabatic influence of the current on the modes is evaluated by considering the phonon retarded Green's function defined as
From Eq.~\ref{eq:langa} we can obtain the nonequilibrium retarded phonon Green's function,
\begin{equation}
D^r(\omega) =  (D^a(\omega))^\dagger = \left [ (\omega+i\eta)^2 - K - \Pi^r(\omega)\right]^{-1} \,,
\end{equation}
and the excitation in terms of the average kinetic energy of the phonons,
\begin{equation}
E_{\rm{kin}}=\int_{-\infty}^{\infty}\frac{d\omega}{2\pi} \omega^2 {\rm{Tr}}\left[D^r(\omega) S_f(\omega) D^a(\omega)\right]\label{eq.Ekin}\,.
\end{equation}
The phonon density of states (DOS) is given by $-2/\pi \omega {\rm Im}\left(D^r(\omega)\right)$.
The DOS is affected both by the coupling to electrons, in particular giving rise to nonequilibrium forces, as well as coupling to the electrode phonons.
In Fig.~\ref{fig:Heating}A we show the phonon DOS at applied bias of $V_a=0$ and $V_a=0.5$V. Most importantly, the DOS becomes negative at a particular phonon frequency ($\omega\approx 170$meV), corresponding to a negatively damped mode, denoted "runaway".
From Eq.~\ref{eq.Ekin} the runaway gives rise to a divergence in the current-induced change of $E_{\rm{kin}}$(heating) of the GNC at $V_a\approx 0.4$V, see Fig.~\ref{fig:Heating}B. This signifies an instability in the harmonic approximation, where the high excitation is likely lead to dramatic effects such as contact disruption\cite{dundas_current-driven_2009}.

The instability can be traced back to the bias dependent electronic friction, and disappears when this is kept at its zero-bias value. We further note that for $V_a$ above $\sim0.3$V the deterministic current-induced forces lead to a qualitatively different heating compared to that of Joule heating only.
Figure~\ref{fig:Heating}B furthermore show how the damping due to electrode phonons is crucial: The heating increase by an order of magnitude if the electrode-phonon bath is neglected. Moreover, if we neglect the damping due to the phonon-bath we observe runaway starting already at $V_a\approx 0.15$V, increasing to more than 15 runaway modes at $V_a\approx 0.4$V, both due to the effects of negative friction and nonconservative forces\cite{dundas_current-driven_2009}. The nonconservative wind and Berry-phase forces are found to be on the same order of magnitude for the runaway mode. Even though they do not themselves lead to the first runaway condition they lower the runaway threshold.

\begin{figure}[htbp]
\centering
%%{\includegraphics[width=0.99\linewidth]{EPSfigures/ModesCombinedV3.eps}}
%{\includegraphics[width=0.99\linewidth]{EPSfigures/ModesCombinedV4.eps}}
%%{\includegraphics[width=0.99\linewidth]{EPSfigures/Qinverse2x2New.eps}}
%{\includegraphics[width=0.99\linewidth]{EPSfigures/Qinverse2x2NewV2b.eps}}
{\includegraphics[width=0.99\linewidth]{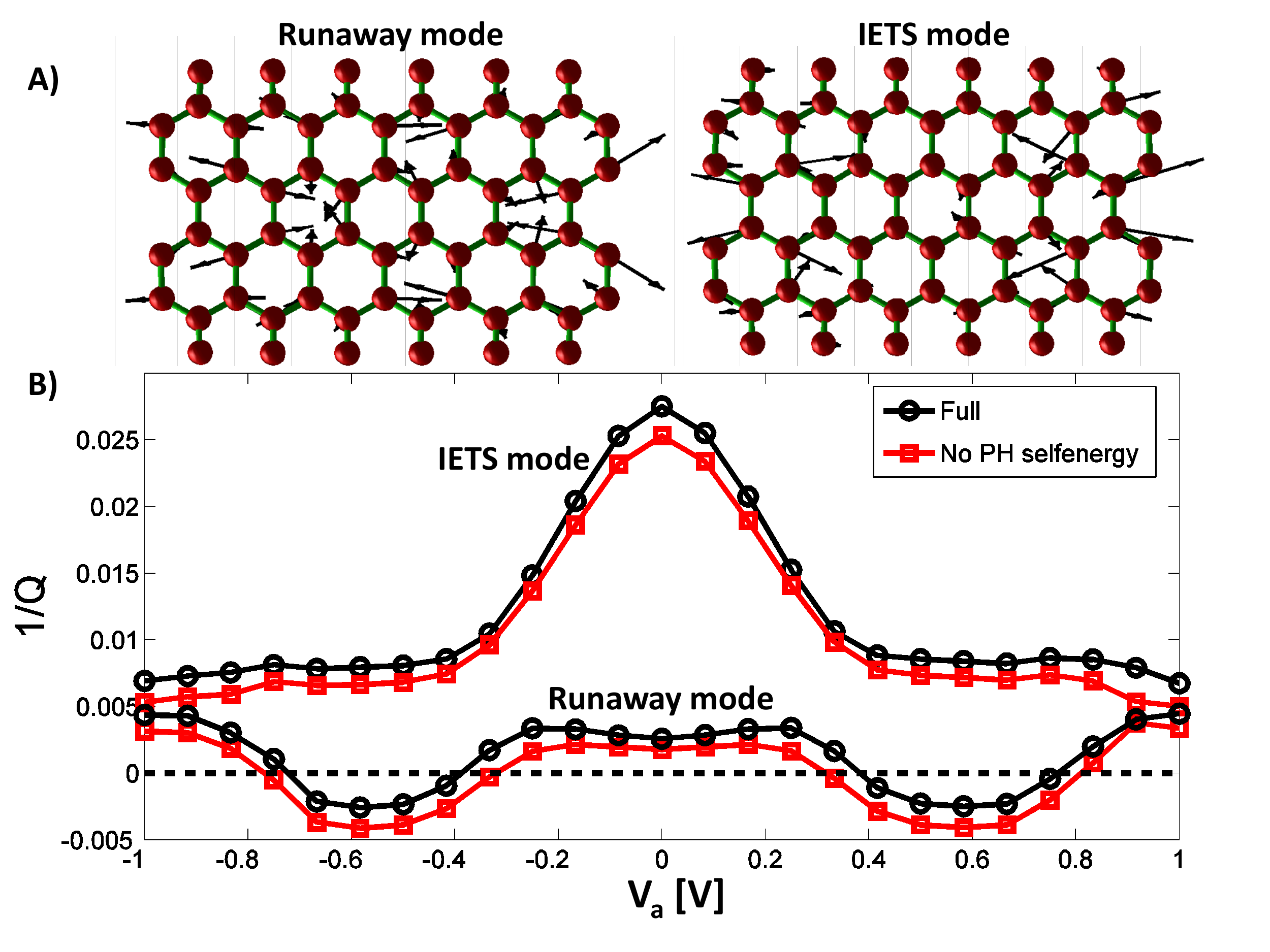}}
\caption{(Color online) A) Two degenerate modes ("runaway"/"IETS") at $V_a=0.4$V with $\omega_0\approx 170$meV. The "runaway" mode break the left-right symmetry due to the coupling to the non-equilibrium electrons and becomes unstable at finite bias. 
The "IETS" yields the largest inelastic signal in the current. B) Inverse $Q$-factor (loss) as a function of bias for the modes.}
\label{fig:Modes}
\end{figure}% (IETS)

We will now in detail analyze the origin of the runaway. We focus on the modes contributing to the phonon DOS peak around the runaway, $\omega_0\approx 170$meV. They can be found as the eigenvectors of $K+{\rm Re}\Pi^r(\omega_0)$.
The two main modes are displayed in Fig.~\ref{fig:Modes}. The "IETS-mode" exhibits the largest inelastic tunnel spectroscopy signal(IETS) in the electronic current and largest noise $S_{f,ii}(\omega_i)$, while the "runaway mode" is the first mode that turns unstable with increasing $V_a$.
In Fig.~\ref{fig:Modes} we show the inverse quality factor, ${1}/{Q} = -2 \frac{{\rm Im}(\omega)}{{\rm Re}(\omega)}=1/Q_{\rm ph}+1/Q_{\rm el}(V_a)$ (energy loss/period), for the two modes as a function of $V_a$. The $Q_{\rm ph}$-factor is relatively big, especially for the runaway mode, due to low phonon DOS around $\omega_0$. The runaway corresponds to amplification, $1/Q<0$, while $1/Q>0$ remains for the IETS mode despite a strong decrease with bias.

%Because the IETS mode is dominated by the intra-electrode contributions its motion is always damped, despite a strong decrease with bias in $1/Q_{\rm el}(V_a)$.
%When the dissipation turns negative, that is $\Im(\omega)<0$, the system has modes characterized by a %diverging amplitude.
%By increasing the bias voltage the nano-bridge will eventually break down much in a similar fashion to %what is known for classical unstable bridge systems. The current-induced forces hereby limit the %current-carrying capacity of GNCs. 
%The observed instability originates from a current-induced negative friction of the runaway mode.

\begin{figure}[htbp]
\centering
%%{\includegraphics[width=0.99\linewidth]{EPSfigures/ABintegrandsV2.eps}}
%%{\includegraphics[width=0.99\linewidth]{EPSfigures/EigChansShifted.eps}}
%%{\includegraphics[width=0.99\linewidth]{EPSfigures/ABcoefficients51V2.eps}}
%{\includegraphics[width=0.99\linewidth]{EPSfigures/TransmissionFrictionRelationV3.eps}}
%{\includegraphics[width=0.99\linewidth]{Fig5}}
{\includegraphics[width=0.99\linewidth]{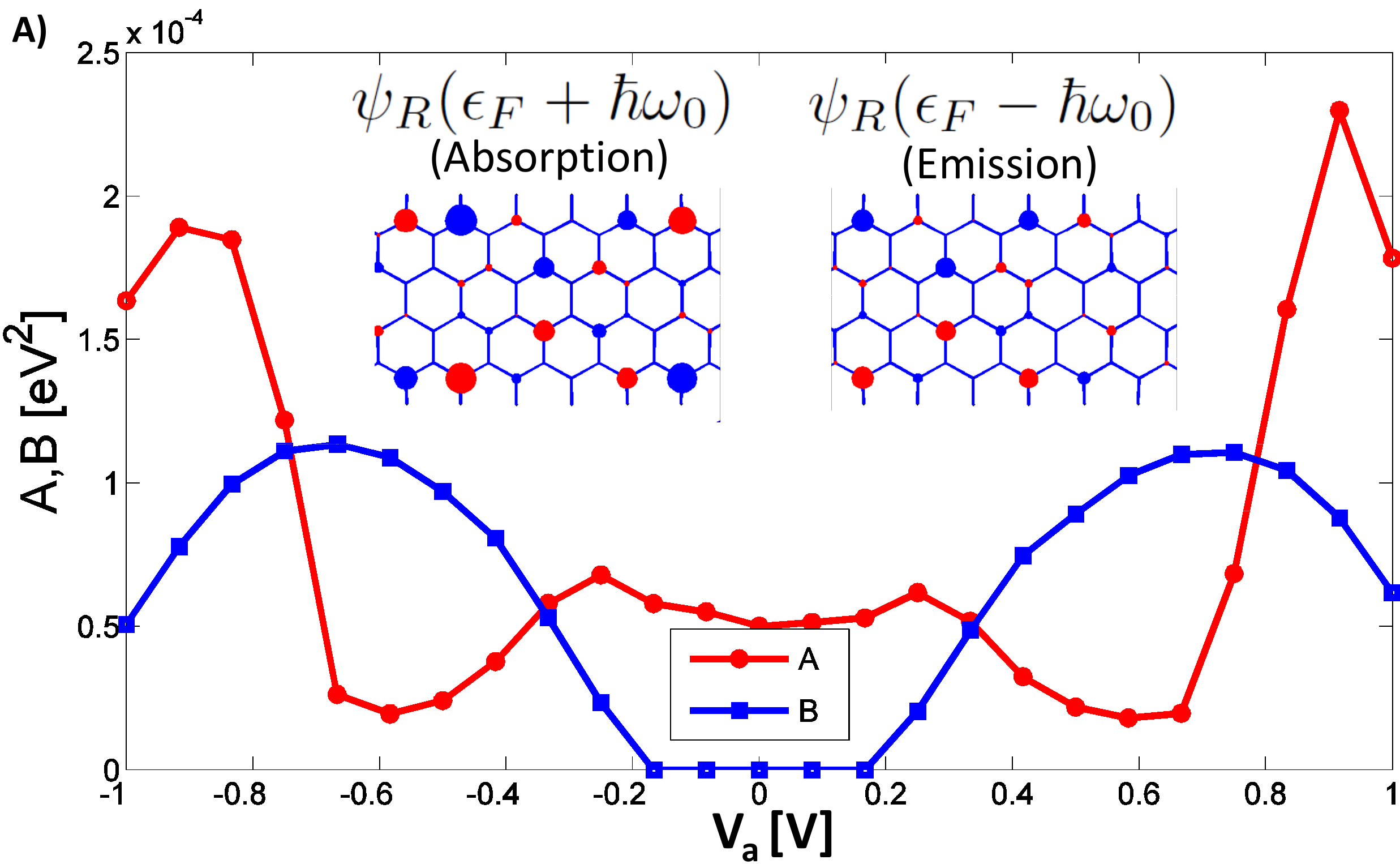}}
{\includegraphics[width=0.99\linewidth]{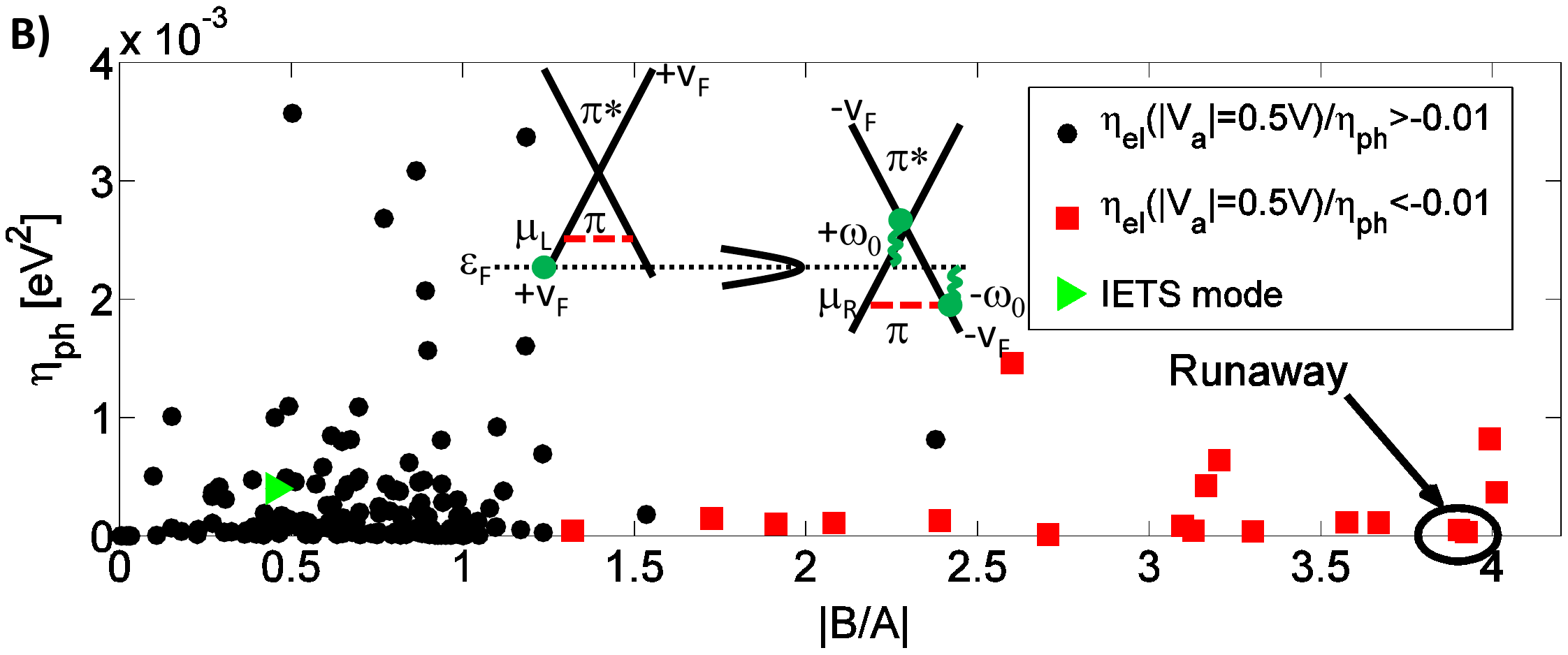}}
\caption{(Color online) Nonequilibrium friction mechanism. A) Phonon absorption/emission ($\mathcal{A}$/$\mathcal{B}$) rates for the runaway mode. Note that $\mathcal{B}=0$ for $V_a<\omega_0$. When $V_a>\pm 0.4$V emission exceeds absorption, $\mathcal{B}>\mathcal{A}$. Insert: At resonance scattering states giving the main contribution to the interaction integrals. The radius shows the absolute value $|\psi(x,y)|$ of the eigenstate, while the color indicates the sign of the real-part. B) Runaway occurs for the mode with the largest emission and lowest phonon friction. Squares indicate modes with a significant electron friction. These modes all have $\mathcal{A}$,$\mathcal{B}$ coefficients with same behavior as the first runaway mode.
Insert: Resonance between two graphene leads at certain filling (red dashed line) and bias voltage. An incoming scattering state (left green dot) at resonance (dashed line) can either absorb ($+\omega_0$) to a state with lower DOS close to the $\pi$-$\pi^*$ crossing or change to a state with higher DOS by emission ($-\omega_0$).}
\label{fig:AB}
\end{figure}

It is instructive to view the runaway in terms of phonon absorption/emission processes in a simple master equation for the phonon number, $N$,
\begin{eqnarray}
    \dot{N} = \mathcal B (N+1) - \mathcal A N,
\end{eqnarray}
where $\mathcal A$($\mathcal B$) are the rates for absorption(emission). 
From Fermi's golden rule we find the emission, 
\begin{eqnarray}
%\mathcal A &=& -2\pi \sum_{\alpha\beta} \left(1+n_B(\hbar\omega+\mu_{\alpha}-\mu_{\beta})\right) \Lambda^{\beta\alpha}(\omega)\,.\\
  &&\mathcal B = -2\pi \sum_{\alpha\beta} n_B(\hbar\omega_0+\mu_{\alpha}-\mu_{\beta}) \Lambda^{\beta\alpha}(\omega_0)\nonumber\,,
\end{eqnarray}
and $\mathcal A$ is obtained by a replacement $\omega\rightarrow -\omega$. Only a single scattering state, $|\psi_{L/R}\rangle$, contributes to $\mathcal A$ and $\mathcal B$. Expressed in the single flux-normalized eigenchannel, and assuming $k_B T\ll\hbar\omega_0 <eV_a$, we have,
\begin{eqnarray}
&&\mathcal B\approx \int_{\mu_R+\hbar\omega_0}^{\mu_L}|\langle \psi_L(\epsilon)|\mathbf{M}|\psi_R(\epsilon-\hbar\omega_0)\rangle|^2 \frac{d\epsilon}{2\pi}\nonumber
\\&&\mathcal A\approx \int_{\mu_R-\hbar\omega_0}^{\mu_L}|\langle \psi_L(\epsilon)|\mathbf{M}|\psi_R(\epsilon+\hbar\omega_0)\rangle|^2 \frac{d\epsilon}{2\pi} \label{eqn:ABapprox}
\end{eqnarray}
Here we did not include the intra-electrode terms($\Lambda^{LL/RR}$) in $\mathcal A$ since these vary only slightly with $V_a$ for the runaway mode.
The phonon absorption rate decrease while the emission rate increase as the bias exceeds the mode frequency, see Fig.~\ref{fig:AB}A.
The electronic friction is given by the difference $\mathcal A - \mathcal B = - 2\pi \sum_{\alpha\beta} \Lambda^{\alpha\beta}(\omega)$. This difference manifests itself in how the $Q$-factor vary with bias for the runaway mode.
%The increased emission does not originate from $a_{L/R}(\epsilon)$, but in how the eigenchannel changes with energy and bias. The interaction matrix element (eqn.~\ref{eqn:ABapprox}) is asymmetric in phonon absorption and emission due to the large mode frequencies. If the main part of the interaction occurs at the lowest chemical potential one can approximate $\mathcal B$ and $\mathcal A$ by $\mathcal B_S \equiv \mathbf{M}^2 T(E_F)(1-T(E_F- \omega_0))\theta(|eV|-\omega_0)$ and $\mathcal A_S \equiv \mathbf{M}^2 T(E_F)(1-T(E_F+\omega_0))$, insert of Fig.~\ref{fig:AB}A. The full interaction is of cause more complicated but the model illustrates the importance of the difference in reflection coefficient at the energies shifted by a phonon frequency.
Importantly, we note that the symmetry of the scattering state $\psi_L^*(E_F)$ is almost unchanged from going {\it up} in energy (absorption), see $\psi_R(E_F + \omega_0)$ shown in the inset in Fig.~\ref{fig:AB}A. On the other hand the symmetry of $\psi_R(E_F - \omega_0)$ differs significantly from this.
Thus we can expect in general that a given phonon will yield very different emission and absorption matrix elements due to the symmetry. 
In particular, the el-ph matrix element of the runaway mode yield very low absorption and high emission due to the selective symmetry of this phonon mode. The large phonon frequencies and linear DOS of graphene strengthens this symmetry breaking.
The negative electronic friction is found for several modes and seems to be a generic phenomena in graphene nanostructures.

In Fig.~\ref{fig:AB}B we illustrate how each mode shows up in a parameter space of the phonon friction and $\mathcal{B}/\mathcal{A}$. The dominating runaway mode shows up at high $\mathcal{B}/\mathcal{A}$ and low phonon friction. The other modes with a non vanishing negative electron friction are also displayed. All these modes have $\mathcal{A}$,$\mathcal{B}$ coefficients with same generic behavior as the first runaway mode (Fig.~\ref{fig:AB}A). In the general case where one has a resonance between graphene leads, insert of Fig.~\ref{fig:AB}B, the wave incoming at resonance will absorb to an eigenstate close to the Dirac crossing. Hence it will have low DOS and a dissimilar phase. On the contrary emission leads to an eigenstate with larger DOS and similar phase. This holds true for states dominated by the inter-lead contributions. Compared to the "runaway" mode the "IETS" mode has low emission-absorption ratio due to high intra-electrode terms, $\Lambda^{LL/RR}$ and a higher phonon damping.

We conclude that negative friction can appear for certain phonons in realistic systems such as graphene nanoconstrictions in the presence of electrical current. The negative friction effect is here rooted in the high phonon energies which lead to markedly different symmetry of the electronic states involved in emission and absorption and thus different matrix elements and rates.  Two-dimensional systems like graphene, where a gate can be applied, makes an exciting test-bed for probing effects of electronic current on the atomic scale.

\emph{Acknowledgments --}
We thank the Danish Center for Scientific Computing (DCSC) for providing computer resources.
The Center for Nanostructured Graphene(CNG) is sponsored by the Danish National Research Foundation.

\end{thebibliography}

% Produces the bibliography via BibTeX.

%\begin{widetext}
%% Wide eqn.
%\begin{equation}
%\label{eq:wideeq}
%\end{equation}
%\end{widetext}

%% Fig:
%\begin{figure}[b]
%\includegraphics{fig_1}% Here is how to import EPS art
%\caption{\label{fig:epsart} A figure caption. The figure captions are
%automatically numbered.}
%\end{figure}

%****************************************************************************************************
%\newpage
%\section{Appendix and further information}

% ar=aspect ratios(width/height): {1:18/21,2:18/14,3:18/18,4:18/14,5:18/19.5}={0.8571,1.2857,1,1.2857,0.9231}
% word count fig: 150/ar+20 = {195,137,170,137,183}=822
% word count: text=1927(hereof 81 in abstract),captions=373,mathlines=9*16/row=144,Inlinemath~102*1/math=102... Total: 1927-81+373+822+144+102=3287.


\begin{thebibliography}{1}
\bibitem [{Hil()}]{Hilbert_def}%
  \BibitemOpen
  \href@noop {} {\ }\bibinfo {note} {The Hilbert transform is defined as
  $\mathcal{H}\{g(x')\}(x) = \frac{1}{\pi}\mathcal{P} \int
  \frac{g(x')}{x'-x}dx'$.}\BibitemShut {Stop}%
\bibitem [{FN_()}]{FN_settings}%
  \BibitemOpen
  \href@noop {} {\ }\bibinfo {note} {We used single-zeta-polarized (SZP)
  localized pseudoatomic orbitals in the basis set for both hydrogen and
  carbon. Exchange-correlation effects were handled within GGA-PBE and a grid
  cutoff of 210 Ry was applied. The Brillouin zone was sampled with a [5 1 2]
  Monkhorst-Pack grid, with the x-axis as the transverse direction and the
  z-axis as the transport direction. The system was relaxed until all forces
  were smaller than 0.005eV/Ang. We used 150 points along the bias contour with
  a broadening of 0.01 eV.}\BibitemShut {Stop}%
\end{thebibliography}
\end{document}